\documentclass[aps,preprint,showkeys,footnote,12pt]{revtex4}
\usepackage{eurosym}
\usepackage{amsfonts}
\usepackage{amsmath}
\usepackage{amssymb}
\usepackage{graphicx}
\usepackage{subfigure}
\usepackage{txfonts}
\usepackage{makeidx}
\usepackage{array,multirow}

\newtheorem{theorem}{Theorem}
\newtheorem{acknowledgement}[theorem]{Acknowledgment}

\input{tcilatex}
\begin{document}

\title{Structural, electronic and topological properties of 3D TmBi compound}
\author{M. Ragragui$^{1,2}$, L. B. Drissi$^{1,2,3,4,\ast }$, S. Lounis$^{4,5}$, E. H. Saidi$%
^{1,2,3} $}
\affiliation{{\small 1-LPHE, Modeling \& Simulations, Faculty of Science, Mohammed V
University in Rabat, Morocco}}
\altaffiliation{{\footnotesize {$\ast $\textbf{Corresponding author:} Lalla Btissam Drissi,
E-mail addresses: lalla-btissam.drissi@fsr.um5.ac.ma, l.drissi@fz-juelich.de}} }
\affiliation{{\small 2- CPM, Centre of Physics and Mathematics, Faculty of Science,
Mohammed V University in Rabat, Morocco }}
\affiliation{{\small 3-College of Physical and Chemical Sciences, Hassan II Academy of
Sciences and Technology, Km 4, Avenue Mohammed VI, Rabat, Morocco }}
\affiliation{{\small 4-Peter Gr\"{u}nberg Institut and Institute for Advanced
Simulation, Forschungszentrum J\"{u}lich \& JARA, Germany}}
\affiliation{{\small 5-Faculty of Physics, University of Duisburg-Essen, 47053 Duisburg,
Germany}}

\begin{abstract}
Using density functional theory based methods we report the structural,
electronic and topological properties of the FCC crystal compound TmBi. This
material is found to be dynamically stable and shows a non magnetic
semimetalic character. By tuning the spin-orbit coupling, we observe a
significant change in the band structure, and the occurrence of band
inversion along $\Gamma -X$ direction. The parity product at time-reversal
invariant momentum points and the Wannier charge center calculations provide
a topological index $Z_{2}=1$ on the $k_{j}=0$ plane (with $j=1,2$ and $3$)
revealing the non trivial topological character of TmBi. The existence of
topologically protected surface states of TmBi through the observation of a
Dirac cones at $\bar{X}$ point confirm our finding. The present work could
inspire platforms for exploring novel topological states within the family
of rare-earth monobismuthides.
\end{abstract}

\maketitle

\section{Introduction}

Topological electronic materials, such as topological insulators (TI) and
topological semimetals (TS) with time reversal (TR) symmetry, are quantum
states of matter exhibiting novel linear responses in the bulk and protected
surface states \cite{A5,A40,A005,A2}. A much richer bulk-boundary
correspondence where gapless states live in boundary corners and hinges of
bulk has also attracted a big deal of attention in the few recent years \cite%
{Ab1,Ab2,Abb2,Ab3,Ab5}. Interestingly, the classification of topological
phases depends on two main ingredients: $(i)$ the dimensionality of the bulk
material and $(ii)$ discrete symmetries protecting boundary states. Refs
\cite{AC1,AC2,AC3} derived first ingredients to capture distinctive features
of topological phases of matter. Other approaches involving the quantized
Wannier center \cite{AD22,AD3}, the signature index \cite{nnous} and the
Berry phase \cite{AADD3} have also been developed.

In conventional topological insulators, the band inversion in the bulk
spectrum, mainly attributed to spin-orbit interaction, is crucial to
distinguish between topological non trivial and topological trivial phases
\cite{AZ1,A3,A46}. Furthermore, the presence of both time reversal and
inversion symmetry in these systems allows the determination of a $%
\mathbb{Z}
_{2}$ topological invariant to confirm the topological non trivial character
of the bulk material \cite{A4,A8,A008}. TIs with broken rotation symmetry
switch to Weyl semimetals \cite{A38}. However, breaking TR symmetry converts
some Dirac semimetals to Weyl semimetals having both zero gap energy at
their Fermi level \cite{A37}. The energy-momentum dispersion is linear in
all momentum directions in Dirac and Weyl semimetals \cite{A28}, and the
corresponding Fermi surfaces are formed by points or lines in topological
nodal semimetals \cite{sem1,sem2,A51}. Topological boundary modes emerging
as chiral Fermi arcs have been observed for Cd$_{3}$As$_{2}$ and Na$_{3}$Bi
using ARPES measurements \cite{A29,A290}. TaAs and WTe$_{2}$ materials are
Type-I Weyl semimetals characterized with standard point-like Fermi surfaces
\cite{A30,A33}.

Rare-earth (RE) monopnictide /monoantimonides/monobismuthides are binary
compounds hosting extremely large magnetoresistence \cite{A52,ndSb,A54}.
Rare-earth monopnictides, including LaX (X =N, P, As, Sb, Bi) and Dy$%
\Upsilon $ ($\Upsilon $=P,As,Sb) compounds exhibit exotic electronic surface
states and different topological character \cite{A49,A53,A56,AAAAO}. A
Dirac-like feature is observed within specific photon energy range in
rare-earth monoantimonides YSb, CeSb, and GdSb, while most RESb (RE=Y, Ce,
Gd, Dy, Ho, Tm, Lu) compounds are topologically trivial \cite{A61,A611}.
ARPES measurements demonstrate possible Dirac fermion in the strongly
correlated cerium monopnictides CeSb, and the absence of band inversion in
both the three dimensional band structure of LaSb and CeSb \cite%
{A55,A580,A5800}.

Despite the increasing interests in RESb/REBi systems and their homologues,
theoretical studies of topological aspects of some rare-earth
monobismuthides are still missing. In this paper, we investigate for the
first time to our knowledge, the face-centred cubic TmBi system with and
without spin orbit coupling (SOC). We first examine its structural stability
using phonon analysis. Then, we show the presence of bulk band inversion and
the existence of surface Dirac cones using different techniques of
topological theory.

The paper is organised as follows; the computational method is detailed in
section 2. We report and discuss the obtained results in Section 3. We end
with a conclusion commenting our findings.

\section{Computational details}

In this article, we employ first principles calculations based on Density
Functional Theory (DFT), with the Projected Augmented Wave (PAW) method
implemented in the Quantum espresso (QE) package \cite{A63}. The generalized
gradient approximation (GGA) is used for the exchange correlation energy
function \cite{A64}. To ensure convergence of our calculations, we employ a
plane-wave basis up to 480 eV and $12\times 12\times 12$ $\Gamma $-centered
K mesh, so that the total energy converges to 1 meV per cell. SOC is
included in the self-consistence electronic structure calculations, and
f-electrons have been treated as core electrons. Since GGA function is known
to exaggerate the band inversion features, we fit the GGA-DFT-band structure
to a tight-binding Hamiltonian using the maximally localized Wannier
function method \cite{A65,A66} with Tm d-orbitals and Bi p-orbitals. Based
on the obtained data, we calculate the hybrid Wannier charge centers(WCCs)
\cite{A67}. These calculations are performed to establish the topological
character using the wanniertools code \cite{A68} using an arbitrary line in
half of Brillouin Zone (BZ) by calculating the number of crossings of WCCs
mod2. The surface spectral functions are also calculated and given. For the
phonon dispersion spectrum, a $4\times 4\times 4$ supercell was used.

\section{Results and discussion}

In this section, we investigate the topological properties of Thulium
monobismuthide using systematic first-principles studies. Optimized lattice
parameters and phonon dispersion spectra with and without SOC are reported.
The presence of bulk band inversion and the existence of surface Dirac cones
are shown.

\subsection{Lattice stability of TmBi}

\begin{figure}[h]
\begin{center}
\includegraphics[scale=0.12]{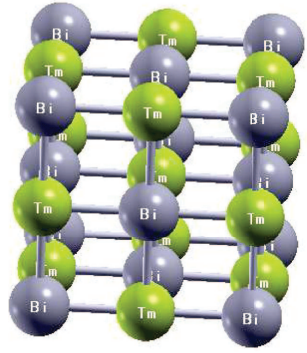}
\end{center}
\caption{{\protect \small {The FCC unit cells of TmBi; the grey balls
correspond to Bi atoms and the yellow ones present the }}${\protect \small Tm}
${\protect \small {\ atoms.}}}
\label{fig1}
\end{figure}

TmBi crystallizes into a face-centred cubic structure belonging to the space
group $Fm\bar{3}m$ (No. 225), in which the Tm atom is positioned at (0.5,
0.5, 0.5) and Bi atom at (0, 0, 0), as shown in Fig.\ref{fig1}. The
optimization of the TmBi structure deals with a lattice parameter a=6.255 $%
\mathring{A}$ used later for the phonons and band structure calculations.

\begin{figure}[h]
\begin{center}
\includegraphics[scale=0.25]{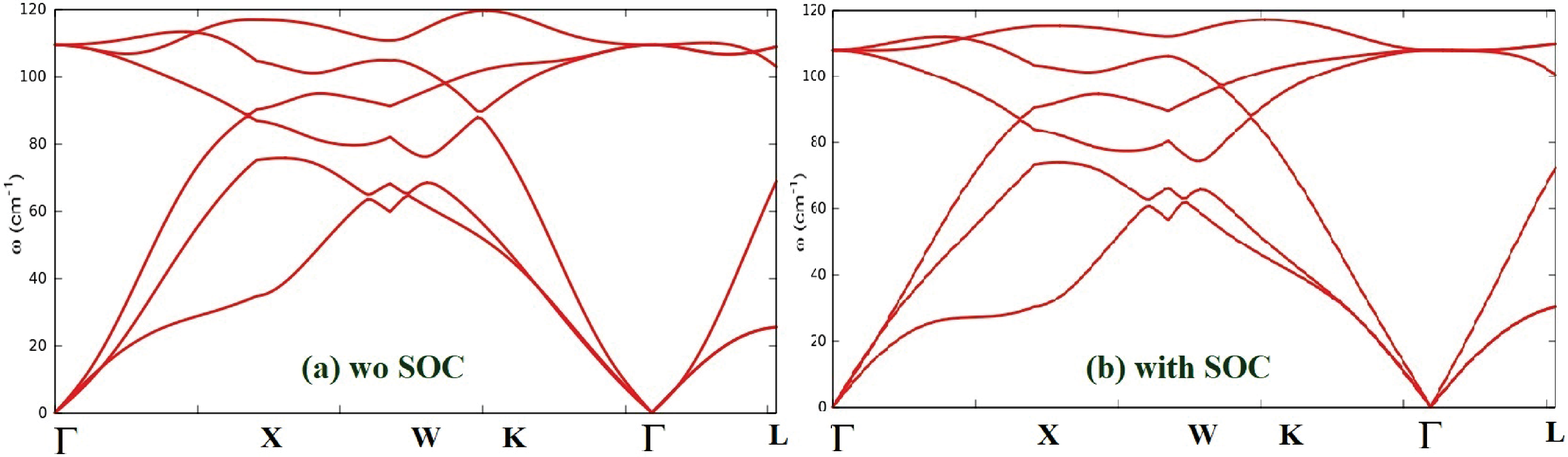}
\caption{{\protect \small {T}mBi's Phonon dispersion spectra: a) without SOC,
b) with SOC.}}
\label{FIG2}
\end{center}
\end{figure}

In the absence of any theoretical and experimental study of lattice dynamics
of TmBi, we investigate the phonon dispersion spectrum with and without
spin-orbit interaction. The shapes of the corresponding curves, plotted in
Figs \ref{FIG2}, are rather similar with a small gap observed between the
acoustic and optical branches in the absence of spin orbit coupling. One can
also notice that the obtained range of phonon frequency is from 0 to 4 $THz$
(120 $cm^{-1}$). Furthermore, no negative frequencies are observed for TmBi
in the Brillouin zone, which indicates a perfect kinetic stability of this
structure and confirms its thermodynamic stability.

\subsection{Electronic properties}

\begin{figure}[h]
\begin{center}
\includegraphics[scale=0.4]{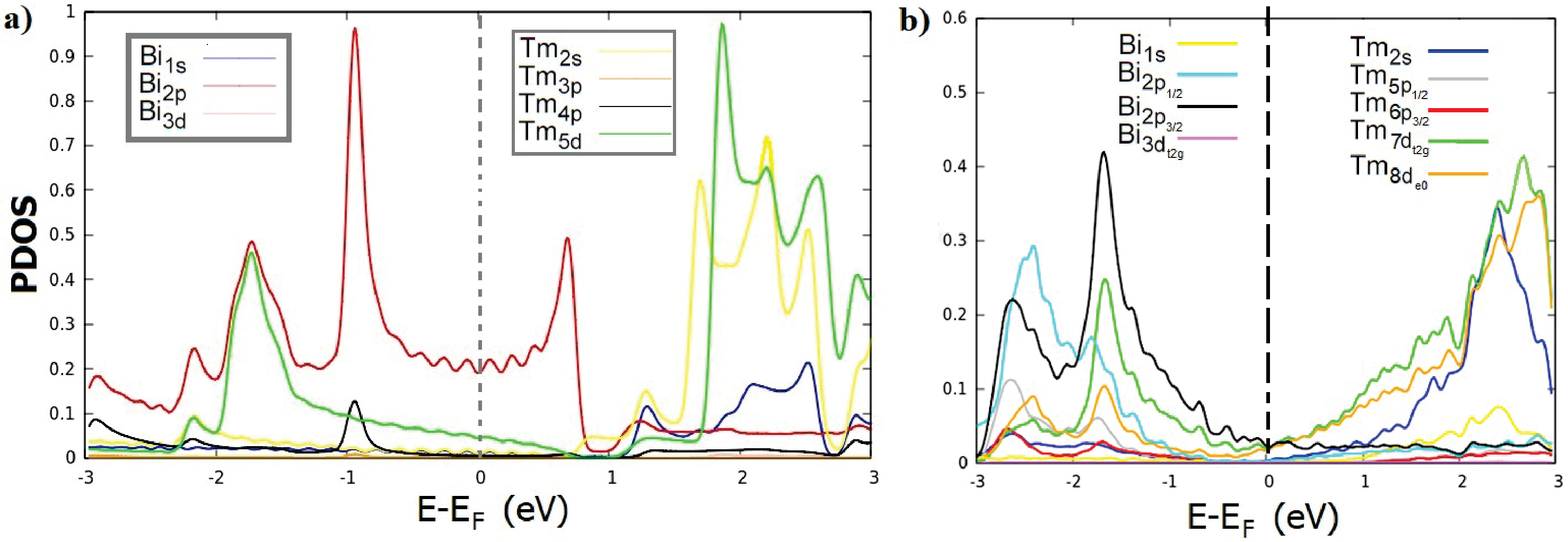}
\end{center}
\caption{{\protect \small Partial density of states (pdos) for a) non-spin-orbit
calculation and b) with SOC.}}
\label{FIG3}
\end{figure}

From the partial density of states (PDOS) curves displayed in Fig.\ref{FIG3}%
, we have analyzed the elementary contribution of all atomic orbitals.
\textbf{In the absence of }SOC, it is shown in Fig.\ref{FIG3}-a that near
the Fermi level, the valence band (VB) is a mixture of Bi(2p)-orbitals and
small contribution of Tm(5d)-orbitals. On the other hand, the conduction
band (CB) mainly consists of Tm(5d) and (2s)-states with a small
contribution of Bi(2p)-orbitals. Strong hybridization between Tm- and
Bi-orbitals is observed in the valence bands far from Fermi level at nearly -%
$1.8$ eV. One deduces that TmBi is a non magnetic semi-metal compound in
good agreement with other topological semimetal materials like PrBi \cite%
{prbi1}, LaBi \cite{A52} and LaSb \cite{A54}.

When SOC the spin-orbit interaction is considered, the partial density of
states in Fig.\ref{FIG3}-b reveals that the Bi p-orbitals splits degenerate
into $\mathbf{p}_{\frac{3}{2}}$- and $\mathbf{p}_{\frac{1}{2}}$- orbitals.
More preciesely, the $\mathbf{p}_{\frac{1}{2}}-Bi$ orbital goes into the
valence band around -2.5 eV and the $\mathbf{p}_{\frac{3}{2}}-Bi$ orbitals
contribute near the Fermi level and in the valence band around -1.5 eV.
Furthermore, the Tm d-orbitals split into e$_{g}$($\mathbf{d}_{z^{2}}$, $%
\mathbf{d}_{x^{2}y^{2}}$) and t$_{2g}$($\mathbf{d}_{xz}$,$\mathbf{d}_{yz}$,$%
\mathbf{d}_{xy}$), where the spin orbit interaction further splits t$_{2g}$
into doubly degenerate and non-degenerate orbitals. Thus, near the Fermi
energy, the bands are dominated by contributions from p-states of Bi and
Tm-d states, whereas other states are mostly unfilled with small
contribution. The present result is comparable with PrBi material \cite{vass}.

\begin{figure}[h]
\begin{center}
\includegraphics[scale=0.25]{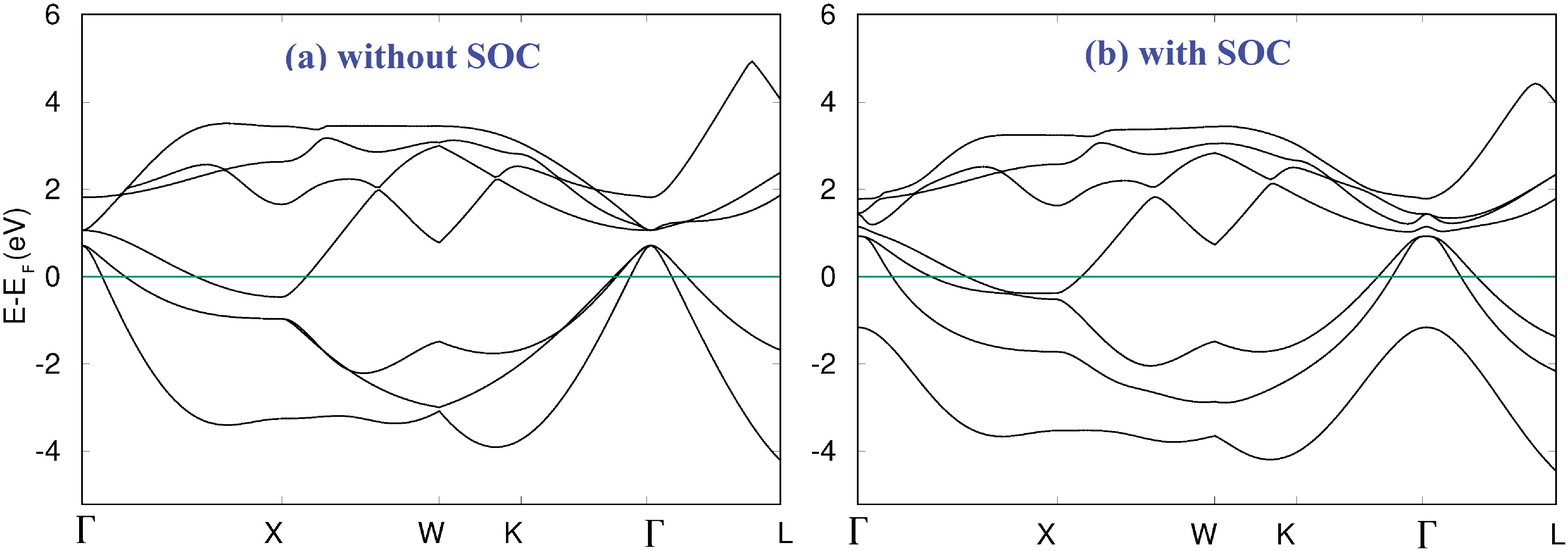}
\end{center}
\caption{{\protect \small Band structure of TmBi (a) without spin orbit
coupling , b) with spin orbit coupling.}}
\label{FIG4}
\end{figure}

The energy band structures of TmBi along high symmetry lines using GGA and
GGA+SOC are plotted in Fig.\ref{FIG4}. In the absence of SOC, there are one
electron-like band at the X-point and three hole-like bands at the $\Gamma -$%
point that cross the Fermi level overlapping together to create a continuous
gap along the whole Brillouin zone in each k-point. Furthermore, a direct
energy gap occurs in two locations: (i) at $0.5$ eV below Fermi level along $%
\Gamma $-X direction and (ii) in the conduction band at $\Gamma $ point.
\newline
When the SOC effect is included into the calculations, a significant change
of the degeneracy of bands-states in the whole Brillouin zone is depicted in
the band structure near $E_{F}$ as shown in Fig.\ref{FIG4}-b. Indeed, the
whole band shifts toward lower binding energy and only a small part of the
band persists above $E_{F}$ at $\Gamma $-point. In turn, the two highest
energy bands below Fermi level become noticeably separated from each other
along $\Gamma $-X path creating a small gap around $43$ meV. The gap
observed at the anticross point increases to $144$ meV at the $X$-point. Our
result within SOC creating an anti-cross is also observed for LaX
(X=As,Sb,Bi) \cite{A51},\cite{A49}-\cite{A56} and reported in earlier
calculations for YsSb \cite{A611} and dysperium monopnictides \cite{AAAAO}.
Furthermore, our calculated band structures with and without SOC, concorde
well with the ones of monobismithudes materials, like LaBi, LaSb, and PrBi
\cite{A5800} since the overall dispersion is very similar, with four bands
crossing the Fermi level, more precisely: threefold degenerate hole-like
bands pinned at about 0.5 eV above E$_{F}$ at the $\Gamma $-point and one
elongated electron band close to the $X$-point at the border of the
Brillouin zone (BZ). It is worth noting that we find the same result using
LDA and LDA+SOC calculations as listed in Table-I. One can see that LDA
slightly increases the gaps with respect to GGA. \newline
This result is a first signature of the topological character of our
compound as confirmed in the the following discussion.
\begin{table}[tbh]
\caption{{\protect \small The energy gap near and at X point with both GGA
and LDA calculations. }}\centering%
\begin{tabular}{c||c|c}
\hline
Gap energy (eV) & Near X point & At X point \\ \hline
GGA & - & 0.498 \\
GGA+SOC & 0.043 & 0.144 \\
LDA & - & 0.4564 \\
LDA+SOC & 0.053 & 0.2 \\ \hline
\end{tabular}%
\end{table}

\begin{figure}[h]
\begin{center}
\includegraphics[scale=0.55]{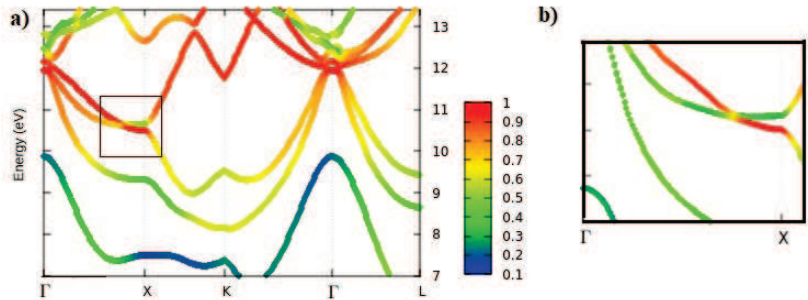}
\end{center}
\caption{{\protect \small Calculated bulk band structure in \ (a) the entire
Brillouin zone and (b) along the }$\Gamma -X${\protect \small \ direction for
TmBi, showing the predicted bulk band inversion between Tm 5d(red) and Bi
2p(green) bands near the X point.}}
\label{FIG5}
\end{figure}

It is clear from Figs.\ref{FIG5} that there is no band inversion at $\Gamma $%
-point. In contrast, the conduction band and valence band give rise to band
inversion along $\Gamma $-$X$ direction, and the orbital character of band
changes near the $X$-point which predict the topological character of TmBi
as reported in other materials such as GdBi and PrBi \cite{A5800}. TmBi
exhibits band inversion between the Tm-5d and Bi-2p states similar to PrBi
\cite{A5800}. Fig.\ref{FIG5} shows the calculated GGA+SOC projected bulk
band structure in the entire Brillouin zone for TmBi. The bands near the
Fermi level mainly come from Tm (5d)-orbitals and Bi (2p)-orbitals, as
indicated by different colors in Fig.\ref{FIG3} and discussed in the PDOS
analysis. Near the $X$-point, the Tm (5d)-band indicated by red color
crosses the green band of Bi (2p), creating an inverted small band gap and
resulting in non-trivial band topology. This band inversion is due to the
odd parity of Bi-2p orbitals and even parity of Tm-5d orbitals. This d-p
band inversion was recently reported for LaBi and Prbi \cite{A56,A5800}, but
such band inversion is absent in LaSb and CeSb \cite{A55}. \newline
In what follows, we examine this topological characteristic according to the
topological theory.

\subsection{Topological index and surface states}

Recall that our system preserves both time-reversal symmetry and inversion
symmetry. It follows that the $%
\mathbb{Z}
_{2}$ topological invariant for the 3D bulk TmBi that characterizes a
topological insulator can be calculated directly by the parity product of
fully occupied valence bands at eight time-reversal invariant momentum
points (TRIM) \cite{A40}. The $\mathbb{Z}_{2}$ topological index, refered to
as $\nu _{0}$, is deduced from the following formula\textrm{\ }%
\begin{equation}
{(-1)}^{\nu _{0}}=\prod_{i=1}^{8}\prod_{n=1}^{N}\delta _{n}(\Gamma _{i})
\label{Z2}
\end{equation}%
where $\delta _{n}$ is the parity product of the valence bands at the $n$-th
TRIM point, the products are over TRIM-point and the Kramers pairs of
occupied bands, respectively\textrm{.} \newline
In this context, our analysis of bands structure of TmBi shown in Fig.\ref%
{FIG4} reveals that there is a gap between conduction and valence bands at
each $k$-point. Within GGA+SOC, we find that TmBi possesses a non trivial
topological index $%
\mathbb{Z}
_{2}=1$. The band parity of the eight TRIM points, namely one $\Gamma $-$%
point,$ three X-points and four L-points, are listed in Table-\ref{tab2}.%
\newline

\begin{figure}[h]
\begin{center}
\includegraphics[scale=0.65]{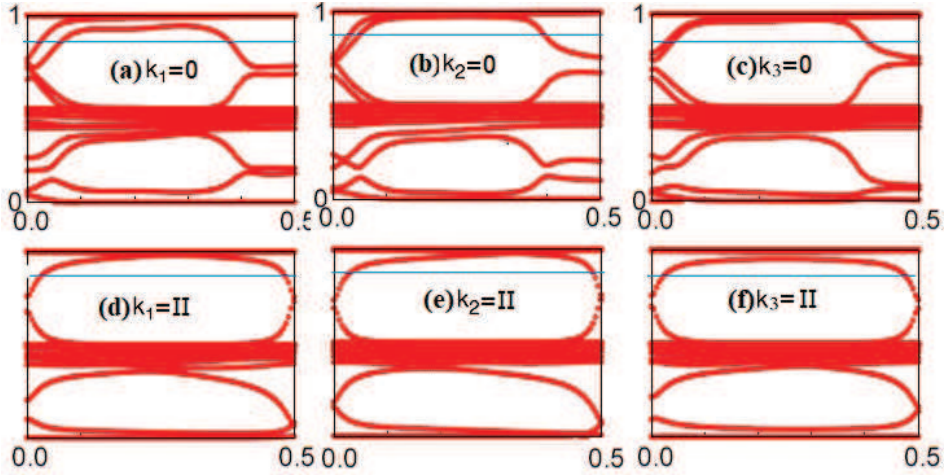}
\end{center}
\caption{{\protect \small Evolution of the Wannier centers }$\protect \theta $%
{\protect \small \ as a function of }$k_{j}${\protect \small \ for the
time-reversal invariant planes (a,b,c) }$k_{j}=0${\protect \small , and
(d,e,f) }$k_{j}=\protect \pi ${\protect \small \ of the bulk Brillouin zone. A
reference line (Blue) is used to determine the }$Z_{2}${\protect \small \
topological invariant of the }$k_{3}-${\protect \small plane.}}
\label{FIG7}
\end{figure}

\begin{table}[tbh]
\caption{{\protect \small Band parity of 8 TRIM points along BZ of TmBi }}
\label{tab2}\centering%
\begin{tabular}{l|l|}
\cline{2-1}
& TmBi \\ \hline
\multicolumn{1}{|l|}{TRIM Points} & $\delta _{n}$ \\ \hline
\multicolumn{1}{|l|}{1$\Gamma $} & + \\ \hline
\multicolumn{1}{|l|}{3X} & - \\ \hline
\multicolumn{1}{|l|}{4L} & + \\ \hline
\end{tabular}%
\end{table}
The $%
\mathbb{Z}
_{2}$-index is also obtained by tracing the evolution of Wannier charge
centers of fully occupied Bloch bands for six time-reversal invariant
momentum planes, namely $\mathrm{(}{\mathrm{k}_{\mathrm{j}}}\mathrm{=0}$ $%
\mathrm{and}$ $\pi \mathrm{,}$ with $j=1,2$ and $3\mathrm{)}$. As clearly
shown in Fig.\ref{FIG7}, the even number of crossings between the reference
line and evolution lines indicates that $%
\mathbb{Z}
_{2}=0$ for $k_{j}=\pi $-plane, while the reference line has odd number of
intersections with the evolution lines confirming that $%
\mathbb{Z}
_{2}=1$ for $k_{j}=0$-plane. Consequently, the $%
\mathbb{Z}
_{2}$ index for TmBi is 1 at (000)-plane which concordes well with our
previous results reported in band structure calculations, predicting the
existence of nontrivial surface states in TmBi. It is worth noting that
similar nontrivial $\mathbb{Z}_{2}$ topological index, and hence topological
surface states with bulk band inversion are predicted, using the same
numerical calculations, for REBi family (with RE ranging from Ce to Gd), for
some rare-earth monobismuthides with partially filled f shell, including
SmSb, GdSb, DyBi, and YbBi, and for dypresium monopnictides compounds,
namely DyP, DyAs and DySb \cite{A55,A56,A5800,AAAAO}.

\begin{figure}[h]
\begin{center}
\includegraphics[scale=0.13]{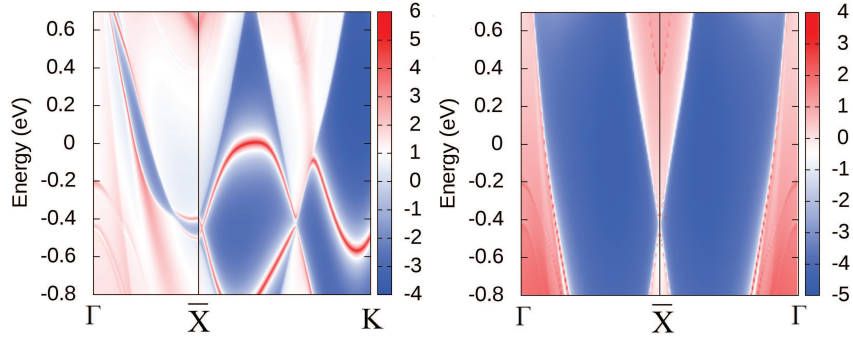}
\end{center}
\caption{{\protect \small Calculated Topological surface states of TmBi along
}$\Gamma -\bar{X}-K${\protect \small \ (a) and along }$\Gamma -\bar{X}-\Gamma
${\protect \small \ (b).}}
\label{FIG8}
\end{figure}

Recall that in the FCC structure rock salt there is three X-points, and one $%
\Gamma $-point. To confirm the topological non-trivial electronic structure
at the $[111]$ surface of TmBi, we firstly perform surface state
calculations along $\Gamma -\bar{X}-M$ direction by employing the Green's
function approach based on a tight-binding model Hamiltonian derived from
our first principles calculations. The result along{\small \ }$\Gamma -\bar{X%
}-K${\small \ }and along{\small \ }$\Gamma -\bar{X}-\Gamma $ is depicted in
Fig.\ref{FIG8} for more visibility. Our calculations reveal that no single
Dirac cone appears at $\Gamma $, while a Dirac cone surface state is clear
observed at $\bar{X}$-point in good agreement with the projection of band
inversion. Therefore, one has an odd number of Dirac cones on the $[111]$
surface. This finding concordes well with LaBi, CeSb, CeBi, PrBi topological
semimetals and Dy$\Upsilon $ compounds with ($\Upsilon $=P,As), where the
presence of topological surface states is prominent \cite%
{A49,A55,A56,A5800}. In contrast, unusual topological surface states
with high anisotropic Dirac-like cone are observed in DySb rock salt
structure and half-Heusler material LnPtBi \cite{A5800,A47}, while only bulk
states are visible in YbSb compound.

It follows that TmBi material can provide an excellent platform for exploring topological surface states in 3D semimetals which are strongly required for applications in spintronics and for electronic devices. Furthermore, it can exhibit high carrier mobility as well as unusual extreme magnetoresistance (XMR) property like some rare-earth monopnictides and monobismithudes previously cited. Therefore, it may help to investigate topologically trivial XMR material and discovery of a relationship between topology and the XMR effect which is still not established.

\section{Conclusion}

Based on density functional theory calculations, the phonon dispersion
spectrum of TmBi compound confirms its structural stability. The bulk
electronic structure of these rock salt-type crystal in the form is
nonmagnetic and semi-metallic. We have examined the low-energy bands near X
point to show significant band inversion due to SOC interactions. To confirm
the topological character of TmBi from the bulk we have determined $%
\mathbb{Z}
_{2}$-invariant for TmBi by two methods, namely the parity product at TRIM
points and the Wannier charge center calculations. Dirac cone surface states
connecting the conduction and valence band, have also been observed. The
present finding on a pristine surface of a strong topological material could
inspire new strategies to design new materials with novel functionalities
and searching for exotic topological non trivial properties.

\begin{acknowledgement}
The authors would like to acknowledge the "Acad\'{e}mie Hassan II des
Sciences et Techniques"-Morocco for its financial support. L. B. Drissi
acknowledges the Alexander von Humboldt Foundation for financial support via
the George Forster Research Fellowship for experienced scientists (Ref 3.4 -
MAR - 1202992).
\end{acknowledgement}

\end{document}